\documentclass[pre,aps,singlecolumn,superscriptaddress,nopacs,floatfix,amssymb,amsmath]{revtex4}
\usepackage{epsfig}
\usepackage{graphics}
\usepackage{color}

\newcommand{\beqn}{\begin{eqnarray}}
\newcommand{\eeqn}{\end{eqnarray}}

\begin{document}

\title{On Solutions to the "Faddeev-Niemi" Equations}

\author{Antti J. Niemi}
\email{Antti.Niemi@physics.uu.se}
\affiliation{Department of Physics and Astronomy, Uppsala University,
P.O. Box 803, S-75108, Uppsala, Sweden}
\affiliation{Laboratoire de Math\'ematiques et Physique Th\'eorique,
Universit\'e Fran\c{c}ois-Rabelais Tours, F\'ed\'eration Denis Poisson - CNRS,
Parc de Grandmont, 37200 Tours, France}
\author{Andrzej Wereszczy\'nski}
\email{wereszcz@th.if.uj.edu.pl}
\affiliation{Institute of Physics, Jagiellonian University, Reymonta 4, Krakow, Poland}

\begin{abstract}
Recently it has been pointed out that the "Faddeev-Niemi" equations that describe the Yang-Mills
equations of motion in terms of a decomposed gauge field, can have solutions that obey the standard
Yang-Mills equations with a source term.  Here we present a general class of such gauge field 
configurations.  They might have physical relevance in a strongly coupled phase, 
where the Yang-Mills theory can not be described in terms of a Landau liquid of  
asymptotically  free gluons. 
\end{abstract}



\maketitle

\section{Introduction}

In the weak coupling limit a Yang-Mills theory describes a  
Landau liquid of weakly interacting, asymptotically free gluons. But it has been proposed by several authors 
that in the strong coupling regime where the physical excitations are different from asymptotically 
free gluons,  it might become more effective to describe the gauge field in a decomposed representation.   
A proper  field decomposition  might also help to identify those field degrees of freedom that  become relevant  
when the theory enters its strongly coupled phase. 
In particular,  the Cho-Duan-Ge \cite{cho}, \cite{duan} decomposition 
describes the gauge field in a manner that directly relates to magnetic monopoles, widely 
presumed to be responsible for the confining properties of the theory.
An on-shell refinement of this decomposition was presented in \cite{fadprl}, where it was 
also shown how the decomposition modifies  the Yang-Mills equations.  In \cite{fadprl}, 
\cite{grav} it was   argued that for {\it generic} field configurations the ensuing 
on-shell decomposed Yang-Mills equations coincide with the conventional Yang-Mills equations.  See however 
\cite{lud1} for an off-shell  completion of the decomposition, and \cite{lud2}, \cite{lud3} for a different 
kind of manifestly complete decomposition based on the concept
of spin-charge separation; for a properly spin-charge separated  gauge field  
the decomposed equations coincide with the original Yang-Mills equations.

In \cite{grav} it was pointed out that besides the generic solutions to the "Faddeev-Niemi" (FN) 
equations in  \cite{fadprl},  there can also be {\it non-generic} 
field configurations. In general these give rise to a source term in  the original 
Yang-Mills equations.  In a recent article \cite{point} it has been pointed out that these non-generic  
field configurations include a constant strength color-electric field that has an obvious attractive physical
appeal. Here we report on a 
more general class of such non-generic solutions of the FN equations. 

\section{Decomposing Yang-Mills}

The decomposed four dimensional $SU(2)$ gauge field introduced in  \cite{fadprl} is
\begin{equation}
\mathbf A_\mu \ = \ C_\mu \mathbf  n + 
\partial_\mu \mathbf n \times \mathbf n +
\rho \,  \partial_\mu \mathbf n  + \sigma\partial_\mu \mathbf n \times \mathbf n
\label{dec1}
\end{equation}
Here $\mathbf n$ is a three component $SU(2)$ Lie-algebra valued 
unit length vector field, $C_\mu$ is a vector  field in $\mathbb R^4$ (we use Euclidean signature) and $\rho,\ \sigma$ are two real
scalar fields.
The reason for the separation between the second and the fourth  term  in (\ref{dec1})  is that it allows us to identify 
the first two terms with the Cho-Duan-Ge connection \cite{cho},
\cite{duan}.

When (\ref{dec1}) is substituted into the Yang-Mills action the ensuing
equations of motion obtained by varying the variables $(C_\mu, \  \mathbf n, \ \rho, \ \sigma)$  yield the following FN equations
\cite{fadprl},
\begin{eqnarray}
\mathbf n \cdot \nabla_\mu \mathbf F_{\mu\nu} \ = \ 0
  \\
 \partial_{\nu} \mathbf  n \cdot \nabla_\mu \mathbf F_{\mu\nu} \ = \ 0 
  \\
\partial_{\nu} \mathbf  n \times \mathbf n  \cdot \nabla_\mu \mathbf F_{\mu\nu} \ = \ 0 
  \\
\left( \nabla_{\nu} \rho + \nabla_{\nu} \sigma \cdot \mathbf  n \times \right)  \nabla_\mu \mathbf F_{\mu\nu} \ = \ 0 
\label{fuleq old}
\end{eqnarray}

In order to get a better understanding of relation between solutions of this set of equations and the original Yang-Mills 
field equations we follow \cite{grav} to introduce a more geometrical framework. Namely, we use a right-handed 
orthonormal triplet  $(\mathbf e_\theta, \mathbf e_\varphi, \mathbf n)$ and define 
\begin{equation}
\kappa_{\mu}^a=\mathbf e_a \cdot \partial_{\mu} \mathbf n
\end{equation}
so that
\[
\mathbf A_{\mu} = C_{\mu} \mathbf n +(\kappa_{\mu}^2 \mathbf e_1 -\kappa_{\mu}^1 \mathbf e_2)  +
+ (\rho \kappa^1_{\mu}+\sigma \kappa^2_{\mu} ) \mathbf e_1+ (\rho \kappa^2_{\mu} - \sigma \kappa^1_{\mu}) \mathbf e_2
\]
In these variables the FN equations read
\cite{grav}
\begin{eqnarray}
\mathbf n \cdot \nabla_\mu \mathbf F_{\mu\nu} \ = \ 0
\nonumber  \\
 \kappa^+_\nu \, \mathbf e_+ \! \cdot \nabla_\mu \mathbf F_{\mu\nu}  \ \equiv \  \kappa^+_\nu (\nabla_\mu \mathbf F_{\mu\nu})^+ \ = \ 0
\nonumber  \\
 \nabla_\nu \phi \,  \mathbf e_- \! \cdot \nabla_\mu \mathbf F_{\mu\nu} \ \equiv \  \nabla_\nu \phi  \, (\nabla_\mu \mathbf F_{\mu\nu})^- \   = \ 0 
\label{fuleq}
\end{eqnarray}
where the corresponding field strength tensor is
\begin{equation}
\mathbf F_{\mu\nu} \ = \  \mathbf n \{ (\partial_\mu C_\nu - \partial_\nu C_\mu) 
- [ 1 - (\rho^2 + \sigma^2)]  \,
\mathbf n \cdot \partial_\mu \mathbf n \times
\partial_\nu \mathbf n \} 
+ \frac{1}{4} \nabla_\mu \phi \left[ \kappa_{\nu}^+ ( \mathbf e^+ + \mathbf e^-) + 
\kappa_\nu^-  ( \mathbf e^+ - \mathbf e^-)  \right] 
- (\mu \leftrightarrow \nu) + c.c.
\label{Ftot}
\end{equation}
with
\[
\kappa^+_{\mu} = \kappa^1_{\mu} + i \kappa^2_{\mu}
\]

In \cite{fadprl}  it was argued  that these equations reproduce the original four dimensional
Yang-Mills equations, for {\it generic}   $\kappa^+_\nu $ and $ \nabla_\nu \phi$. Subsequently 
this was shown to be the case in two dimensions (coordinates $x_1=x$, $x_2=y$) \cite{grav}, 
where using antisymmetry of $\mathbf F_{\mu\nu}$  it was shown that 
the last two equations in (\ref{fuleq}) can be written as the following homogeneous linear system,
\begin{equation}
{\mathcal M^\alpha}_\beta (\nabla \,  \mathbf F_{yx})^\beta \ \equiv \ 
\equiv \left( \begin{matrix} 
\kappa_x^1 & - \kappa_y^1 & -\kappa_x^2 & \kappa_y^2 \\
\kappa_x^2 & - \kappa_y^2 & \kappa_x^1 & -\kappa_y^1 \\
\nabla_x \rho & -\nabla_y \rho & \nabla_x \sigma & - \nabla_y \sigma \\
\nabla_x \sigma & -\nabla_y \sigma & - \nabla_x \rho & \nabla_y \rho \end{matrix} \right) 
\left( \begin{matrix} 
 (\nabla_y \mathbf F_{yx})^1 \\ (\nabla_x \mathbf F_{yx})^1 \\
  (\nabla_y \mathbf F_{yx})^2 \\ (\nabla_x \mathbf F_{yx})^2
   \end{matrix} \right)
\ = \ 0
\label{mat4}
\end{equation}
Consequently, for {\it generic} two dimensional field configurations that is field configurations 
for which the determinant of the $4 \times 4$ matrix  $\mathcal M$ in ({\ref{mat4})
does not vanish, the FN  equations (\ref{fuleq}) reproduce the original Yang-Mills equations \cite{grav}
\[
\nabla_\mu \mathbf F_{\mu\nu} \ = \  0
\]
Therefore, one may expect that for {\it non-generic} field configurations leading to the vanishing determinant in (\ref{mat4}),
solutions of the FN 
equations may possibly not obey the original Yang-Mills  equations. Indeed, 
recently \cite{point} have investigated solutions of (2)-(5)  that do not obey the original four dimensional Yang-Mills equations.
As an Ansatz the authors considered essentially two dimensional ({\it e.g.} $x_3$ and $x_4$ independent) gauge  fields in $\mathbb R^4$. 
The ensuing solutions of (\ref{fuleq}) that fail to satisfy the original Yang-Mills equations in $\mathbb R^4$ are then obtained by
looking for such two dimensional decomposed  gauge fields (\ref{dec1}) for which 
the determinant of the $4\times 4$  matrix $\mathcal M$  in (\ref{mat4}) vanishes: 
In  the explicit Ansatz in \cite{point} the elements on the second and fourth columns of this matrix  are all zero.
This leads to a constant strength color-electric Yang-Mills field, a solution of the original Yang-Mills equations in the presence 
of an external source \cite{point}.  Since a constant strength color-electric field has an obvious physical appeal, it is worth while
to study further the properties of the non-generic solutions of the FN equations.

\section{Solving the FN equations}

We shall now show  that there are additional familiar  and physically appealing 
field configurations  that  solve the FN equation but are described by the original Yang-Mills
equations with a source term.  In particular,  it appears that many known classical
solutions of Yang-Mills theory that give rise to a source term, are {\it sourceless} solutions of (\ref{fuleq}).

Since the structure of (\ref{fuleq}) is relatively
simple, we expect that its  non-generic solutions  can be described in 
quite general terms. Here we look only for configurations that appear as solutions to
\begin{equation}
\nabla_\mu \phi = 0
\label{solu1}
\end{equation}
For these field configurations only the $\mathbf n$-component of (\ref{Ftot}) survives,
\begin{equation}
\mathbf F_{\mu\nu} \ \to  \  \mathbf n \,  G_{\mu\nu} 
\equiv \ \mathbf  n \{ (\partial_\mu C_\nu - \partial_\nu C_\mu) 
- [ 1 - |\phi|^2]  \,
\mathbf n \cdot \partial_\mu \mathbf n \times
\partial_\nu \mathbf n \}  
\equiv \mathbf n \{ F_{\mu \nu} - (1-|\phi|^2) H_{\mu \nu}\}
\label{Fn}
\end{equation}
where
\begin{eqnarray}
F_{\mu \nu} &=& \partial_\mu C_\nu - \partial_\nu C_\mu, \\ 
H_{\mu \nu} &=& \mathbf n \cdot \partial_\mu \mathbf n \times
\partial_\nu \mathbf n \\
G_{\mu\nu} &=& F_{\mu\nu} + H_{\mu\nu}
\end{eqnarray}
Let us now analyze how (\ref{solu1}) impacts on the FN equations. Obviously, the third equation in (\ref{fuleq}) is identically fulfilled. The second equation leads to 
\begin{eqnarray}
 \rho (\kappa^1_{\nu}\kappa^2_{\mu} - \kappa^2_{\nu} \kappa^1_{\mu}) G_{\mu \nu} =0, \\ 
 (1+\sigma) (\kappa^1_{\nu}\kappa^2_{\mu} - \kappa^2_{\nu} \kappa^1_{\mu}) G_{\mu \nu} =0
\end{eqnarray}
which for nontrivial case i.e., $\rho \neq 0$ and $\sigma \neq -1$ is equivalent to an orthogonality condition   
\begin{equation}
H_{\mu \nu} G_{\mu \nu} =0 \label{constrain}
\end{equation}
where we use 
\[
H_{\mu \nu} =  \kappa^1_{\mu}\kappa^2_{\nu} - \kappa^2_{\mu} \kappa^1_{\nu}
\] 
Finally, the first equation in (\ref{fuleq}) gives
\begin{equation}
\partial_{\mu} G_{\mu \nu}=0 \label{mx}
\end{equation}
To summarize, the FN equations in the sector defined by (\ref{solu1}) are equivalent to the Maxwell equations for $G_{\mu \nu}$ (\ref{mx}) and constrain (\ref{constrain}).

\vskip 0.2cm
Consider the full Yang-Mills equations for the choice (\ref{solu1})
\begin{equation}
\nabla_{\mu} \mathbf F_{\mu \nu} = \partial_{\mu} \mathbf F_{\mu \nu} + \mathbf A_{\mu} \times\mathbf F_{\mu \nu} 
=  \partial_{\mu} \mathbf n \;  G_{\mu \nu} +\mathbf n  \; \partial_{\mu}  G_{\mu \nu}  + \mathbf A_{\mu} \times \mathbf n \; G_{\mu \nu} 
\end{equation} 
However, assuming that fields obey the FN equations we get
\begin{equation}
\nabla_{\mu} \mathbf F_{\mu \nu} = \mathbf J_{\nu} 
\end{equation} 
where we find the following generally non-vanishing external current
\begin{equation}
\mathbf J_{\nu} = \left( \rho \; \partial_{\mu} \mathbf n \times \mathbf n - \sigma \;  \partial_{\mu} \mathbf n\right)  G_{\mu \nu}  
\end{equation}
Now, we are able to present several examples of sourceless configurations of the FN equations which are solutions to the Yang-Mills equations with the above  source term. In the simplest case we assume $$H_{\mu \nu} \equiv 0$$ which identically solves the constraint. Although this tensor identically vanishes, the unit vector field $\mathbf n$ does not need to be trivial. It may for example simply depend {\it arbitrarily} on one  single space-time coordinate, lets say $\mathbf n = \mathbf n (x^{\lambda})$, where $\lambda$ is a fixed index. Thus, we are left with $U(1)$ gauge theory for $C_{\mu}$  $$ \partial_{\mu} F_{\mu \nu}=0$$
Solution discussed in \cite{point} belongs to this class., it can be easily generalized to a configuration for which the field tensor is independent of the 
coordinate $F_{\mu \nu}= const$. Namely, 
\[
C_{\mu} = a_{\mu} + b_{\mu \nu} x_{\nu}
\] 
where $b_{\mu \nu}$ is an arbitrary four dimensional constant matrix. Then 
\[
F_{\mu \nu} = b_{\nu\mu} - b_{\mu \nu}
\] 
and the external current reads
\begin{equation}
\mathbf J_{\nu} = \left( \rho \; \partial_{\lambda} \mathbf n \times \mathbf n - \sigma \;  \partial_{\lambda} \mathbf n\right) (b_{\nu\lambda} - b_{\lambda \nu})
\end{equation}  
where no summation on $\lambda$ is assumed. 
\\
The constancy of the field tensor is by no means essential to our construction. For example, one can consider the plane wave solution propagating along $z$-axis with frequency $\omega$ 
\begin{eqnarray*}
C_0=C_z=0 \\ 
C_1=a \sin \omega (z-t) \\
C_2=a\cos \omega (z-t)
\end{eqnarray*}
It solves the sourceless FN equations whereas for the full Yang-Mills equation there is a source
\begin{equation}
\mathbf J_{\nu} = a \omega \left( \rho \; \partial_x \mathbf n \times \mathbf n - \sigma \;  \partial_x \mathbf n\right) \cos \omega (z-t) \cdot \left( \begin{matrix}
-1 ~ \\ 0 \\ 0 \\1 \end{matrix} \right) 
\end{equation}

Observe that the external current vanishes if the so-called valence degrees of freedom are absent $\rho=\sigma=0$ \cite{cho}. 
This may indicate a particularly
close 
relation in $(2+1)$ dimensions  between the FN and Yang-Mills equations, due to the fact that in (2+1) dimensional case the nonabelian gauge field has 3 on-shell degrees of freedom. Then, it is sufficient to use the Cho-Duan connection containing only the U(1) field $C_{\mu}$ (one field degree of
freedom) and $\mathbf n$ (two field degrees of freedom).

\section{Conclusions}
In conclusion, we have generalized the observation made  in \cite{point},  that the "Faddeev-Niemi" equations have physically appealing solutions that solve
the original Yang-Mills equation with a source term. The formalism that we have presented allows a more systematic analysis of such solutions. Since 
the decomposed representation of the Yang-Mills field is presumed to identify those excitations that become important in strongly coupled 
phases of the Yang-Mills theory that can not be described in terms of asymptotically free gluons and conventional weak coupling perturbation 
theory, it should be of interest to better understand the relevance of the solutions of the "Faddeev-Niemi" equations to the non-perturbative structure of Yang-Mills theories.

\begin{acknowledgments}
We thank Maxim Chernodub for comments.
\end{acknowledgments}

\end{document}